\documentclass[rmp,aps,preprint,nofootinbib,endfloats*,a4paper,notitlepage]{revtex4}

\usepackage{amsmath}
\usepackage{amssymb}
\usepackage{amsfonts}
\usepackage{epsfig, color, ulem}
\usepackage{graphicx}
\usepackage{tcolorbox}

\usepackage{natbib}
\usepackage{setspace}
\newlength{\dinwidth}
\newlength{\dinmargin}
\usepackage{lineno}

\begin{document}

\begin{spacing}{2}

\title{On the relation between the propagator matrix and the Marchenko focusing function}
\author{\small Kees Wapenaar$^1$ and Sjoerd de Ridder$^2$}
\affiliation{$^1$Delft University of Technology, Department of Geoscience and Engineering, Stevinweg 1, 2628 CN Delft, The Netherlands\\
$^2$University of Leeds, School of Earth and Environment, Leeds, LS2 9JT, United Kingdom\\
\mbox{}\\
{\rm Right-running head: Propagator matrix and focusing function}}

\date{\today}

\begin{abstract}
The propagator matrix ``propagates'' a full wave field from one depth level to another, accounting for all propagation angles and evanescent waves. 
The Marchenko focusing function forms the nucleus of data-driven Marchenko redatuming and imaging schemes, accounting for internal multiples. 
These seemingly different concepts appear to be closely related to each other.
With this insight, the strong aspects of the propagator matrix (such as the handling of evanescent waves) can be transferred to the focusing function.
Vice-versa, the propagator matrix inherits from the focusing function that it can be retrieved from the reflection response,
which reduces its sensitivity to the subsurface model.
\end{abstract}
\maketitle

\section{Introduction}

The propagator matrix \citep{Gilbert66GEO, Kennett72GJRAS, Woodhouse74GJR} ``propagates'' a wave field from one depth level to another. It acts on the full wave field and hence
it implicitly accounts for downgoing and upgoing, propagating and evanescent waves. Unlike one-way propagation operators used in seismic migration, the propagator matrix does not
depend on the square-root operator. This facilitates its numerical implementation, particularly for waves with large propagation angles.
\citet{Kosloff83GEO} proposed to use 
the propagator matrix concept in seismic migration and called this ``migration with the full acoustic wave equation.''
They used filters to eliminate evanescent and downward propagating waves, hence, they only exploited the advantageous numerical aspects.
\citet{Wapenaar86GP2} exploited the fact that the propagator matrix (which they called the ``two-way wavefield extrapolation operator'') 
simultaneously handles downgoing and upgoing waves and proposed a migration scheme that accounts for internal multiples.
In this method the propagator matrix is defined on basis of a detailed subsurface model.
Because this method appeared to be very sensitive to the used model, it has not been developed beyond horizontally-layered medium applications.

The Marchenko method has been introduced as a data-driven way to deal with internal multiples in seismic redatuming and imaging (Wapenaar et al., 2014; Broggini et al., 2014).
It uses focusing functions that are retrieved from the reflection response at the surface and a macro velocity 
model that only needs to explain the direct arrival of the focusing functions.
The Marchenko method is in principle suited to handle internal multiples in large-scale 3D imaging problems 
(Pereira et al., 2019; Staring and Wapenaar, 2020; Ravasi and Vasconcelos, 2021).

\citet{Becker2016SEG}, \citet{Wapenaar2017GP2} and \citet{Elison2020PHD} indicate how full wavefield propagation methods
\cite{Kosloff83GEO, Wapenaar93GJI} can be used to model the Marchenko focusing function when a detailed subsurface model is available.
Here we present a more general discussion on the relation between the propagator matrix and the focusing function and briefly indicate new research directions.

Underlying assumptions of the Marchenko method are that the wave field inside the medium 
can be decomposed into downgoing and upgoing waves and that evanescent waves can be ignored.
Only recently several approaches have been proposed that aim to circumvent these assumptions \cite{Diekmann2021PRR, Kiraz2021JASA, Wapenaar2021GJI}.
In the current paper we show that the Marchenko focusing function can be explicitly expressed in terms of the propagator matrix and vice versa.
On the one hand this allows to extend the validity of the focusing function to full (non-decomposed) wave fields, including evanescent waves.
On the other hand it opens the way to use the propagator matrix in imaging problems, without the usual sensitivity to the subsurface model, because the multiples in the
propagator matrix are now retrieved from the reflection response.

In this paper we limit ourselves to establishing the relation between the propagator matrix and the Marchenko focusing function. 
A detailed discussion of its potential applications is beyond the scope of this paper.

\section{The propagator matrix}

Our starting point is the following matrix-vector wave equation in the space-frequency $({\bf x},\omega)$ domain
\begin{eqnarray}\label{eq2.1}
\partial_3{{\bf q}} = {{{\mbox{\boldmath ${\cal A}$}}}}\,{{\bf q}}+{{\bf d}},
\end{eqnarray}
with wavefield vector ${{\bf q}}({\bf x},\omega)$,   operator matrix ${{{\mbox{\boldmath ${\cal A}$}}}}({\bf x},\omega)$ and source vector ${{\bf d}}({\bf x},\omega)$ defined as
\begin{eqnarray}
{{\bf q}}=\begin{pmatrix} p \\ v_3 \end{pmatrix},\,
{{{\mbox{\boldmath ${\cal A}$}}}}= \begin{pmatrix}0      &i\omega \rho \\
i\omega\kappa-\frac{1}{i\omega}\partial_\alpha\frac{1}{\rho}\partial_\alpha &0    \end{pmatrix},\,
{{\bf d}}=\begin{pmatrix}\hat f_3 \\ q \end{pmatrix}
\label{eq9996ge}
\end{eqnarray}
(Corones, 1975; Kosloff and Baysal, 1983; Fishman and McCoy, 1984; Wapenaar and Berkhout, 1986).
Here $p({\bf x},\omega)$ and $v_3({\bf x},\omega)$ are the pressure and vertical particle velocity of the acoustic wave field, 
  $\kappa({\bf x})$ and $\rho({\bf x})$ the compressibility and mass density of the lossless inhomogeneous medium,
 and $q({\bf x},\omega)$ and $\hat f_3({\bf x},\omega)$ the volume injection rate and external vertical force densities
 (the hat  is used to distinguish the external force from a focusing function).
 Furthermore, $i$ is the imaginary unit 
 and the summation convention holds for repeated subscripts, with Greek subscripts taking the values 1 and 2 only.
The propagator matrix ${\bf W}({\bf x},{{\bf x}_R},\omega)$ is defined as the solution of the source-free wave equation
\begin{eqnarray}\label{eq2.1gw}
\partial_3{\bf W} ={{{\mbox{\boldmath ${\cal A}$}}}}{\bf W},
\end{eqnarray}
with boundary condition
\begin{eqnarray}
{\bf W}({\bf x},{{\bf x}_R},\omega)|_{x_3=x_{3,R}} = {\bf I}\delta({{\bf x}_{{\rm H}}}-{{\bf x}_{{\rm H},R}}),\label{eq9998d}
\end{eqnarray}
with horizontal coordinate vectors ${{\bf x}_{{\rm H}}}=(x_1,x_2)$ and ${{\bf x}_{{\rm H},R}}=(x_{1,R},x_{2,R})$, and ${\bf I}$ denoting a $2\times 2$ identity matrix. 
Let ${{\partial{\mathbb{D}}}}_R$ denote a horizontal boundary at $x_3=x_{3,R}$. 
The propagator matrix ``propagates'' the field ${{\bf q}}({\bf x},\omega)$ from ${{\partial{\mathbb{D}}}}_R$ to any depth level $x_3$ as follows
\begin{eqnarray}
{{\bf q}}({\bf x},\omega)=\int_{{{\partial{\mathbb{D}}}}_R} {\bf W}({\bf x},{\bf x}_R,\omega){{\bf q}}({\bf x}_R,\omega){\rm d}{\bf x}_R\label{eq1330}
\end{eqnarray}
\citep{Gilbert66GEO, Kennett72GJRAS, Woodhouse74GJR}, assuming the source vector ${{\bf d}}$ is zero between ${{\partial{\mathbb{D}}}}_R$ and depth level $x_3$.
We partition ${\bf W}$ as follows
\begin{eqnarray}\label{eq424}
{\bf W}({\bf x},{\bf x}_R,\omega)= \begin{pmatrix}W^{p,p}      & W^{p,v} \\
                W^{v,p} & W^{v,v}    \end{pmatrix}({\bf x},{\bf x}_R,\omega),
\end{eqnarray}
with the first and second superscript referring to the field quantities at ${\bf x}$ and ${\bf x}_R$, respectively. 
From equations \ref{eq2.1gw} and \ref{eq9998d} and the structure of ${{{\mbox{\boldmath ${\cal A}$}}}}$ in equation \ref{eq9996ge} it follows that
$W^{p,p}$ and $W^{v,v}$ are real-valued, whereas  $W^{p,v}$ and $W^{v,p}$ are imaginary-valued. 
The propagator matrix can be built up recursively, according to
\begin{eqnarray}\label{eq65awc}
{\bf W}({\bf x},{\bf x}_R,\omega)=\int_{{{\partial{\mathbb{D}}}}_A}{\bf W}({\bf x},{\bf x}_A,\omega){\bf W}({\bf x}_A,{\bf x}_R,\omega){\rm d}{\bf x}_A,
\end{eqnarray}
where ${{\partial{\mathbb{D}}}}_A$ is a horizontal boundary at $x_{3,A}$. The arrangement of $x_{3,R}$, $x_{3,A}$ and $x_3$ is arbitrary.

\begin{figure}
\vspace{-.4cm}
\centerline{\epsfysize=5.5 cm \epsfbox{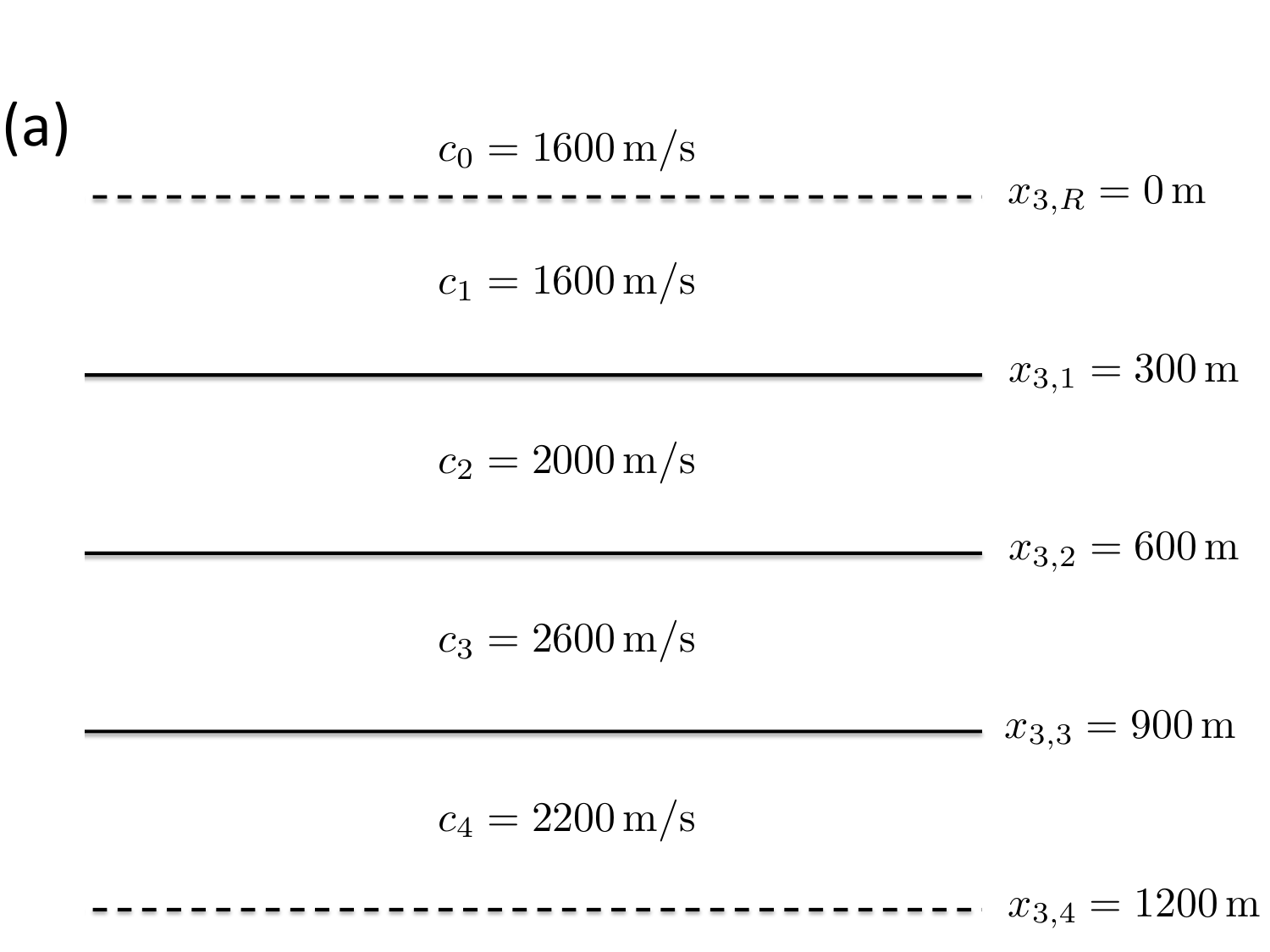}}
\vspace{.3cm}\centerline{\epsfysize=5.5 cm \epsfbox{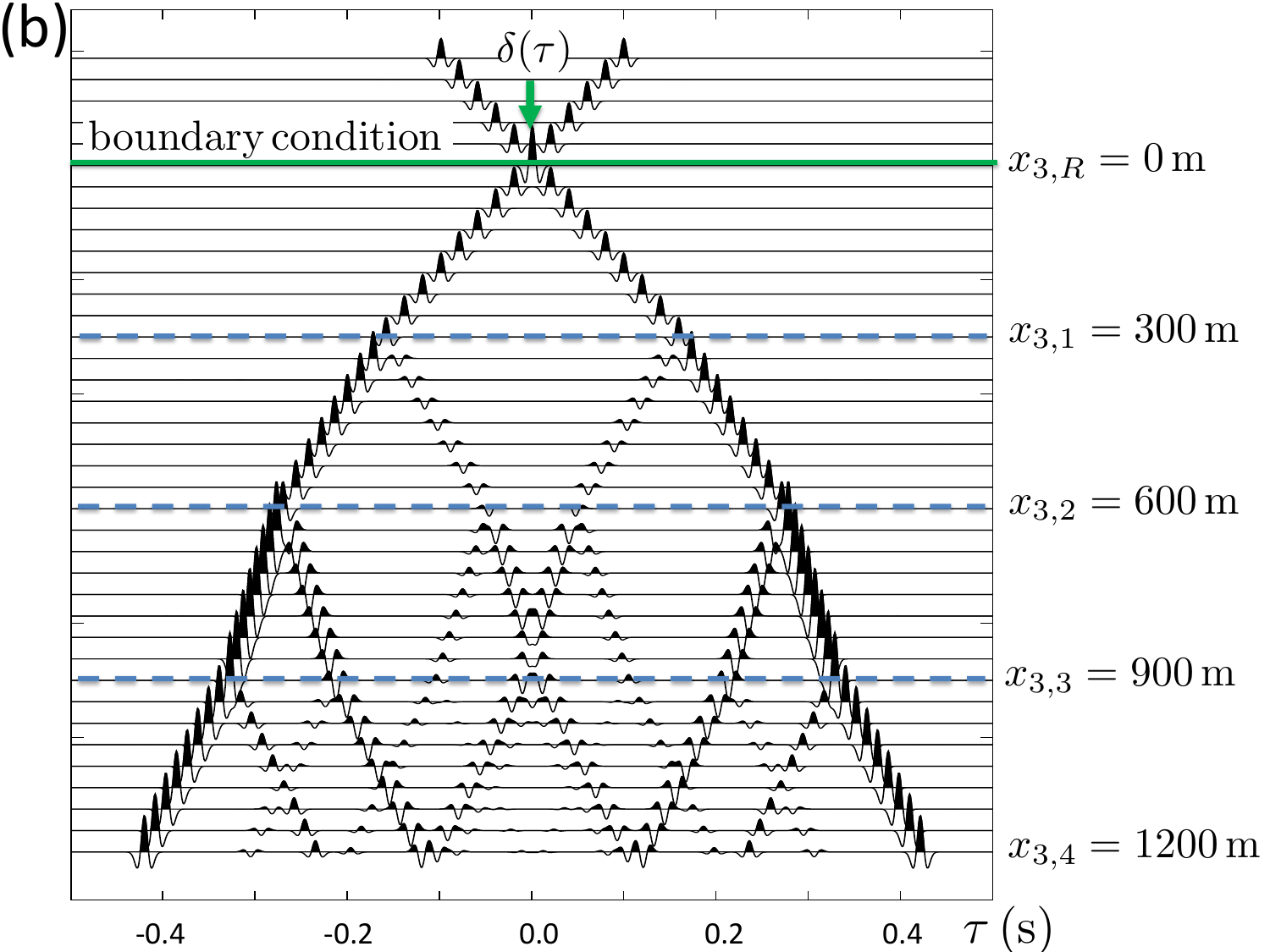}}
\vspace{-0cm}
\vspace{0cm}\centerline{\epsfysize=5.5 cm \epsfbox{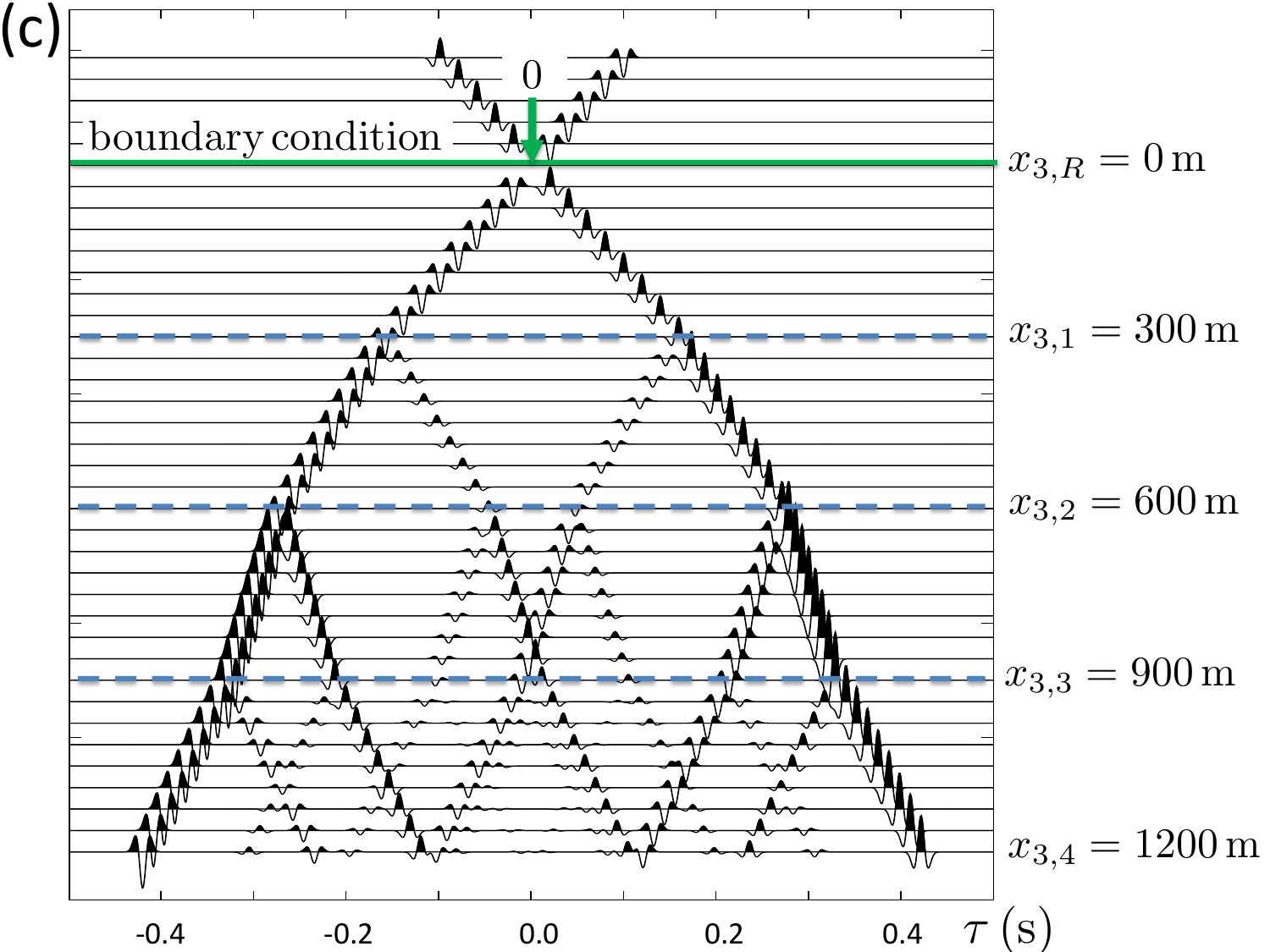}}
\vspace{-0cm}
\centerline{\epsfysize=5.5 cm \epsfbox{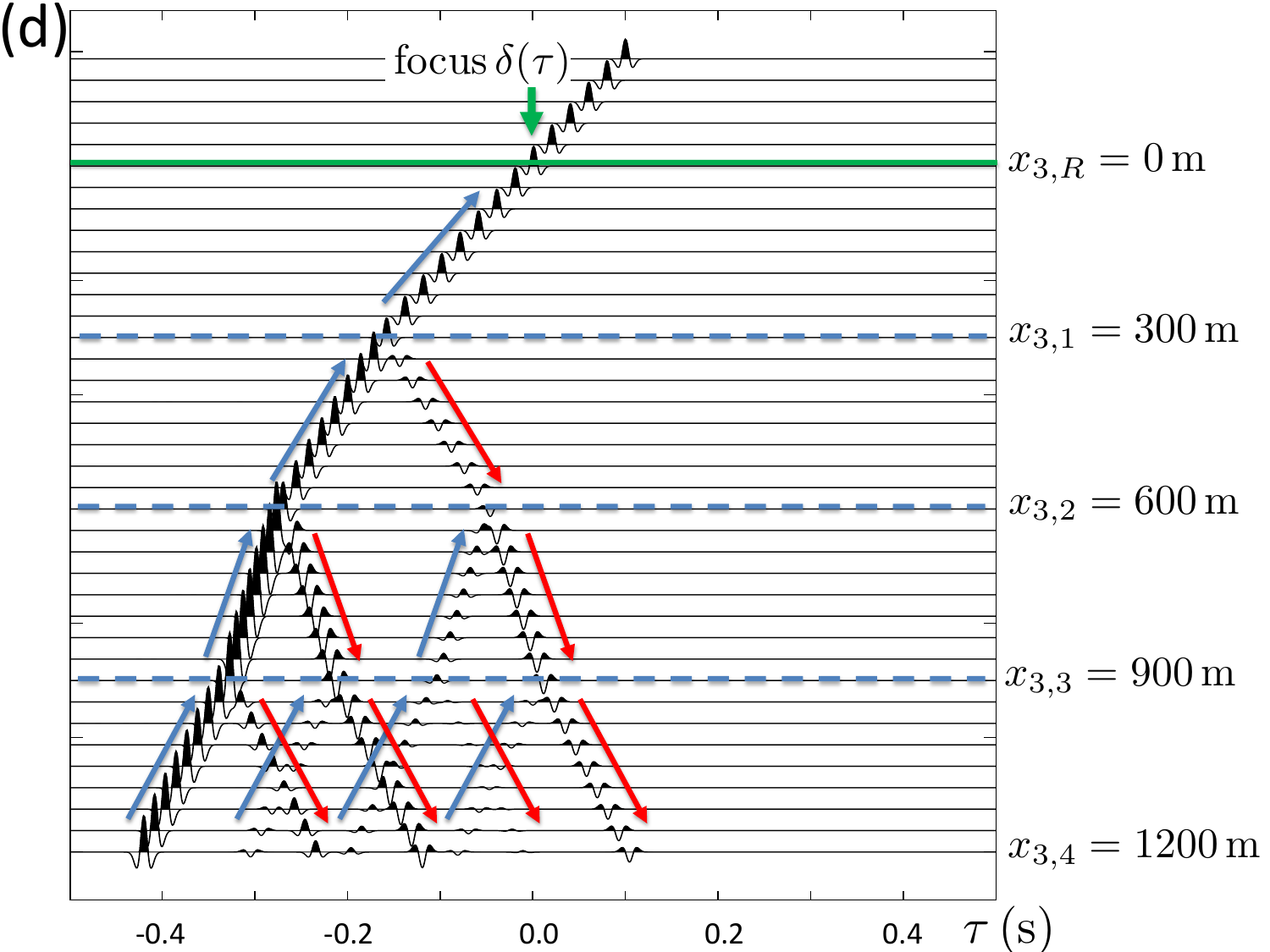}}
\caption{\small (a) Horizontally layered medium. 
(b) Symmetric propagator element $W^{p,p}(s_1,x_3,{x_{3,R}},\tau)$  (fixed $s_1$ and $x_{3,R}$), convolved with a wavelet. 
(c) Anti-symmetric propagator element  $W^{p,v}(s_1,x_3,{x_{3,R}},\tau)$. (d) Focusing function $F(s_1,x_3,{x_{3,R}},\tau)$.}\label{Figure1}
\end{figure}

As an illustration, we consider the propagator matrix for a laterally invariant medium. For this situation 
it is convenient to consider the propagator matrix  in the horizontal slowness domain,
i.e., $\tilde{\bf W}(s_1,x_3,x_{3,R},\omega)$, with $s_1$ denoting the horizontal slowness. In a homogeneous layer, the elements of  $\tilde{\bf W}$ are given by
\begin{eqnarray}
\tilde W^{p,p}(s_1,x_3,x_{3,R},\omega)&=&\cos(\omega s_3\Delta x_3),\\
\tilde W^{p,v}(s_1,x_3,x_{3,R},\omega)&=&\frac{i\rho}{s_3}\sin(\omega s_3\Delta x_3),
\end{eqnarray}
$\tilde W^{v,p}=(s_3^2/\rho^2)\tilde W^{p,v}$ and $\tilde W^{v,v}=\tilde W^{p,p}$, with $\Delta x_3=x_3-x_{3,R}$, and vertical slowness $s_3$ defined as  $s_3=\sqrt{1/c^2-s_1^2}$, 
with propagation velocity $c=1/\sqrt{\kappa\rho}$.
These expressions hold for propagating and evanescent waves. For propagating waves ($s_1^2\le 1/c^2$), their temporal inverse Fourier transforms read 
\begin{eqnarray}
&&W^{p,p}(s_1,x_3,x_{3,R},\tau)={\frac{1}{2}}\{\delta(\tau-s_3\Delta x_3)+\delta(\tau+s_3\Delta x_3),\label{eq412}\\
&&W^{p,v}(s_1,x_3,x_{3,R},\tau)=\frac{\rho}{2s_3}\{\delta(\tau-s_3\Delta x_3)-\delta(\tau+s_3\Delta x_3), 
\end{eqnarray}
etc., where $\tau$ is the intercept time. 
Note that $W^{p,p}$ and $W^{v,v}$ are symmetric, whereas $W^{p,v}$ and $W^{v,p}$ are anti-symmetric.
For the horizontally layered medium of Figure 1a, Figure 1b 
shows the symmetric element $W^{p,p}(s_1,x_3,{x_{3,R}},\tau)$  as a function of $x_3$ and $\tau$, convolved with a Ricker wavelet with a central frequency of 50 Hz,
 for a single horizontal slowness $s_1=1/3000$ s/m.
The trace at $x_3=x_{3,R}=0$ m shows the boundary condition $W^{p,p}(s_1,x_{3,R},x_{3,R},\tau)=\delta(\tau)$.
The traces between $x_{3,R}$ and $x_{3,1}$ show the two delta functions in the right-hand side of equation \ref{eq412} (convolved with the Ricker wavelet).
The traces in the deeper layers are the result of the recursive application of equation \ref{eq65awc}
in the slowness intercept-time domain.
In a similar way, Figure 1c shows the anti-symmetric element
$W^{p,v}(s_1,x_3,{x_{3,R}},\tau)$. The trace at $x_3=x_{3,R}=0$ m shows the boundary condition $W^{p,v}(s_1,x_{3,R},x_{3,R},\tau)=0$. 

\section{The Marchenko focusing function}

From here onward we let ${{\partial{\mathbb{D}}}}_R$ at depth $x_{3,R}$ denote a transparent acquisition boundary. 
The medium above this boundary is homogeneous; below this boundary the medium is inhomogeneous and source-free.
Before we return to the 3D situation, we discuss the Marchenko focusing function $F(s_1,x_3,x_{3,R},\tau)$ for a horizontally layered medium in the slowness intercept-time domain.
This focusing function is a solution of the wave equation, with focusing condition $F(s_1,x_{3,R},x_{3,R},\tau)=\delta(\tau)$. Hence, $F$ focuses at the acquisition boundary, similar as
the focusing function $f_2$ of \citet{Wapenaar2014GEO}. 
Further we demand that $F$ is purely upgoing at and above ${{\partial{\mathbb{D}}}}_R$. This focusing function is illustrated in
Figure 1d for the horizontally layered medium of Figure 1a.
At the bottom we see four upgoing waves (indicated  by the blue arrows), which are tuned such that at $x_3=x_{3,R}=0$ a single upgoing wave focuses at $\tau=0$.
Note that $F$ in Figure 1d resembles a number of events of the propagator element $W^{p,p}$ in Figure 1b. As a matter of fact, 
 $F$ can be expressed as a combination of the symmetric and  anti-symmetric functions  $W^{p,p}$  and  $W^{p,v}$ of Figures 1b and 1c, according to
\begin{eqnarray}
&&F(s_1,x_3,{x_{3,R}},\tau)=
W^{p,p}(s_1,x_3,{x_{3,R}},\tau) - \frac{s_{3,0}}{\rho_0}W^{p,v}(s_1,x_3,{x_{3,R}},\tau),\label{eq1135}
\end{eqnarray}
where $\rho_0$ and $s_{3,0}$ are the mass density and vertical slowness, respectively, of the homogeneous upper half-space. 
Conversely, using the fact that $W^{p,p}$ is symmetric and $W^{p,v}$ is anti-symmetric, we can construct these elements from the 
focusing function $F$, according to 
\begin{eqnarray}
&&W^{p,p}(s_1,x_3,{x_{3,R}},\tau)=
{\frac{1}{2}}\{F(s_1,x_3,{x_{3,R}},\tau) +F(s_1,x_3,{x_{3,R}},-\tau)\},\label{eq1135vv}\\
&&W^{p,v}(s_1,x_3,{x_{3,R}},\tau)=
-\frac{\rho_0}{2s_{3,0}}\{F(s_1,x_3,{x_{3,R}},\tau) -F(s_1,x_3,{x_{3,R}},-\tau)\}.\label{eq1136vv}
\end{eqnarray}
We now return to the 3D situation and derive relations similar to equations \ref{eq1135} $-$ \ref{eq1136vv}.
In the homogeneous upper half-space (including the boundary ${{\partial{\mathbb{D}}}}_R$) we define pressure-normalized downgoing and upgoing waves $p^+$ and $p^-$, respectively.
In the space-frequency domain, we relate these fields to $p$ and $v_3$ via 
${\bf q}={{{\mbox{\boldmath ${\cal L}$}}}}{\bf p}$, with ${{\bf q}}$ defined in equation \ref{eq9996ge} and
\begin{eqnarray}
{{{\mbox{\boldmath ${\cal L}$}}}}=\begin{pmatrix}1 & 1\\ \frac{1}{\omega\rho_0}{\cal H}_1 & -  \frac{1}{\omega\rho_0}{\cal H}_1\end{pmatrix},\quad
{\bf p}=\begin{pmatrix}p^+ \\p^-\end{pmatrix}.\label{eq517}
\end{eqnarray}
Here ${\cal H}_1$ is the square-root of the Helmholtz operator $\omega^2/c_0^2+\partial_\alpha\partial_\alpha$ in the homogeneous upper half-space
\citep{Corones75JMAA, Fishman84JMP, Wapenaar86GP2}.
Substitution of ${\bf q}={{{\mbox{\boldmath ${\cal L}$}}}}{\bf p}$ into equation \ref{eq1330} gives,
for ${\bf x}$ in the inhomogeneous and source-free half-space below ${{\partial{\mathbb{D}}}}_R$,
\begin{eqnarray}
{{\bf q}}({\bf x},\omega)=\int_{{{\partial{\mathbb{D}}}}_R} {\bf Y}({\bf x},{\bf x}_R,\omega){\bf p}({\bf x}_R,\omega){\rm d}{\bf x}_R,\label{eq1330dec}
\end{eqnarray}
for $x_3\ge x_{3,R}$, with ${\bf Y}({\bf x},{\bf x}_R,\omega)={\bf W}({\bf x},{\bf x}_R,\omega){{{\mbox{\boldmath ${\cal L}$}}}}({\bf x}_R,\omega)$.
From equation \ref{eq1330dec}, using equations  6 and 15,  we obtain for the first element of vector ${{\bf q}}$
\begin{eqnarray}
p({\bf x},\omega)&=&\int_{{{\partial{\mathbb{D}}}}_R} F^*({\bf x},{\bf x}_R,\omega)p^+({\bf x}_R,\omega){\rm d}{\bf x}_R 
+\int_{{{\partial{\mathbb{D}}}}_R} F({\bf x},{\bf x}_R,\omega)p^-({\bf x}_R,\omega){\rm d}{\bf x}_R, \label{eq12}
\end{eqnarray}
for $x_3\ge x_{3,R}$ (superscript $*$ denotes complex conjugation), with focusing function $F({\bf x},{\bf x}_R,\omega)$ defined as
\begin{eqnarray}
&&F({\bf x},{\bf x}_R,\omega)=W^{p,p}({\bf x},{\bf x}_R,\omega)- \frac{1}{\omega\rho_0}{\cal H}_1({\bf x}_R,\omega)W^{p,v}({\bf x},{\bf x}_R,\omega).\label{eqFdef}
\end{eqnarray}
Here we used the fact that $W^{p,p}$ and $W^{p,v}$ are real- and imaginary-valued, respectively. 
Moreover, we used that ${\cal H}_1({\bf x}_R,\omega)$ is a symmetric operator and assumed it is real-valued, which implies that we ignored evanescent waves at ${{\partial{\mathbb{D}}}}_R$.
Equation \ref{eq12} was derived previously via another route \cite{Wapenaar2021GJI}; the explicit expression for $F$ in equation \ref{eqFdef} is new.
Note the analogy with the definition of the focusing function  in the slowness intercept-time domain in equation \ref{eq1135}.
From equations  \ref{eq9998d}, \ref{eq424} and \ref{eqFdef} we find
$F({\bf x},{\bf x}_R,\omega)|_{x_3=x_{3,R}} = \delta({{\bf x}_{{\rm H}}}-{{\bf x}_{{\rm H},R}})$, which confirms that $F$ is indeed a focusing function. The focusing function is visualised in Figure 2a.
Conversely, using  that $W^{p,p}$ and $W^{p,v}$ are real- and imaginary-valued, respectively, we  find
\begin{eqnarray}
&&W^{p,p}({\bf x},{\bf x}_R,\omega)=\Re\{F({\bf x},{\bf x}_R,\omega)\},\label{eq523}\\
&&W^{p,v}({\bf x},{\bf x}_R,\omega)=-i\omega\rho_0{\cal H}_1^{-1}({\bf x}_R,\omega)\Im\{F({\bf x},{\bf x}_R,\omega)\},\label{eq524}
\end{eqnarray}
where $\Re$ and $\Im$ stand for the real and imaginary part, respectively.
From equations  \ref{eq2.1gw} and \ref{eq424}, with ${{{\mbox{\boldmath ${\cal A}$}}}}$ defined in equation \ref{eq9996ge}, we obtain for the other two elements of the propagator matrix
\begin{eqnarray}
W^{v,p}({\bf x},{\bf x}_R,\omega) &=& \frac{1}{i\omega\rho({\bf x})}\partial_3W^{p,p}({\bf x},{\bf x}_R,\omega),\label{eq1126W}\\
W^{v,v}({\bf x},{\bf x}_R,\omega) &=& \frac{1}{i\omega\rho({\bf x})}\partial_3W^{p,v}({\bf x},{\bf x}_R,\omega).\label{eq1124W}
\end{eqnarray}

\begin{figure}
\centerline{\epsfysize=7.5 cm \epsfbox{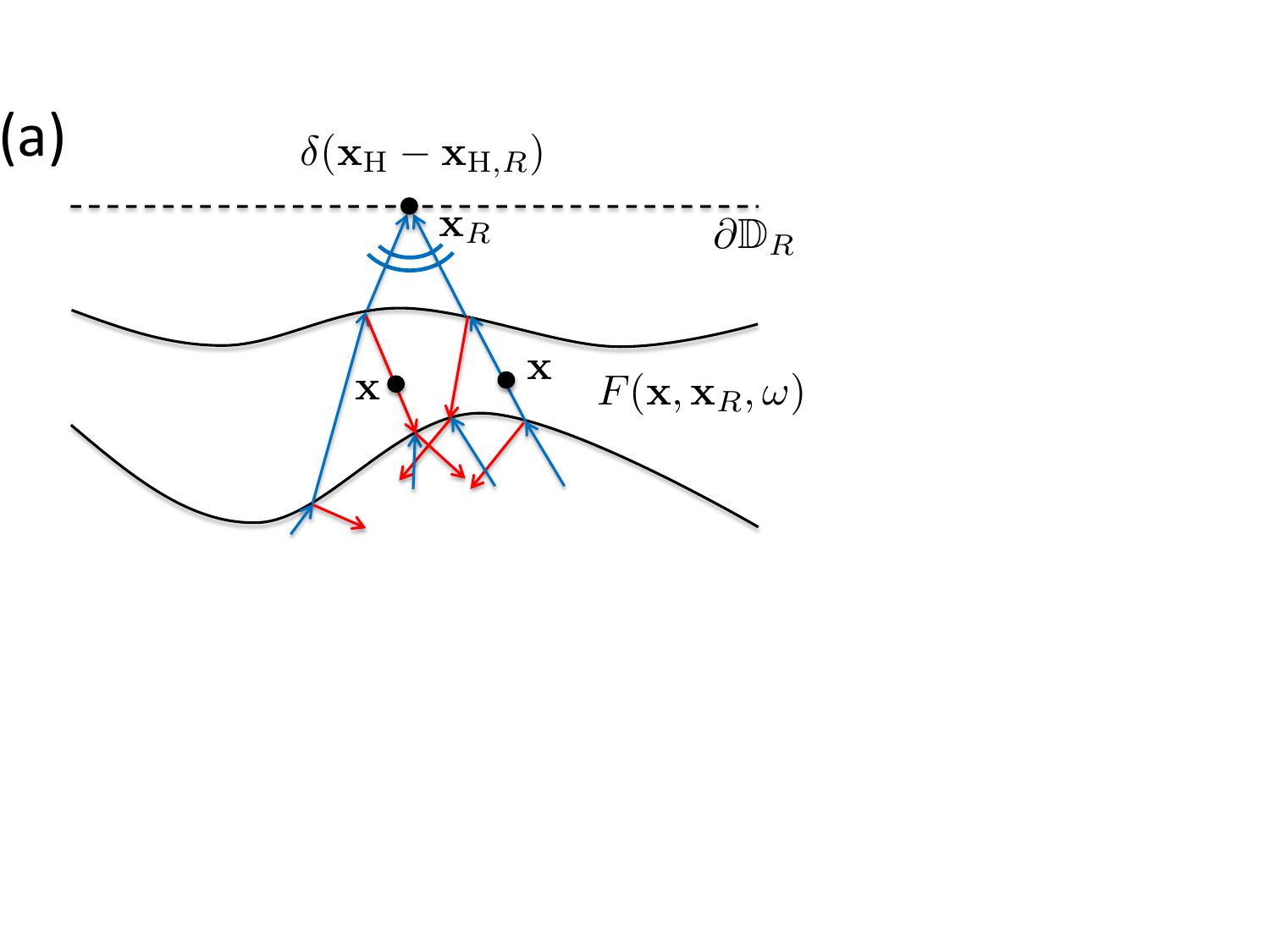}}
\vspace{-1cm}\centerline{\epsfysize=7.5 cm \epsfbox{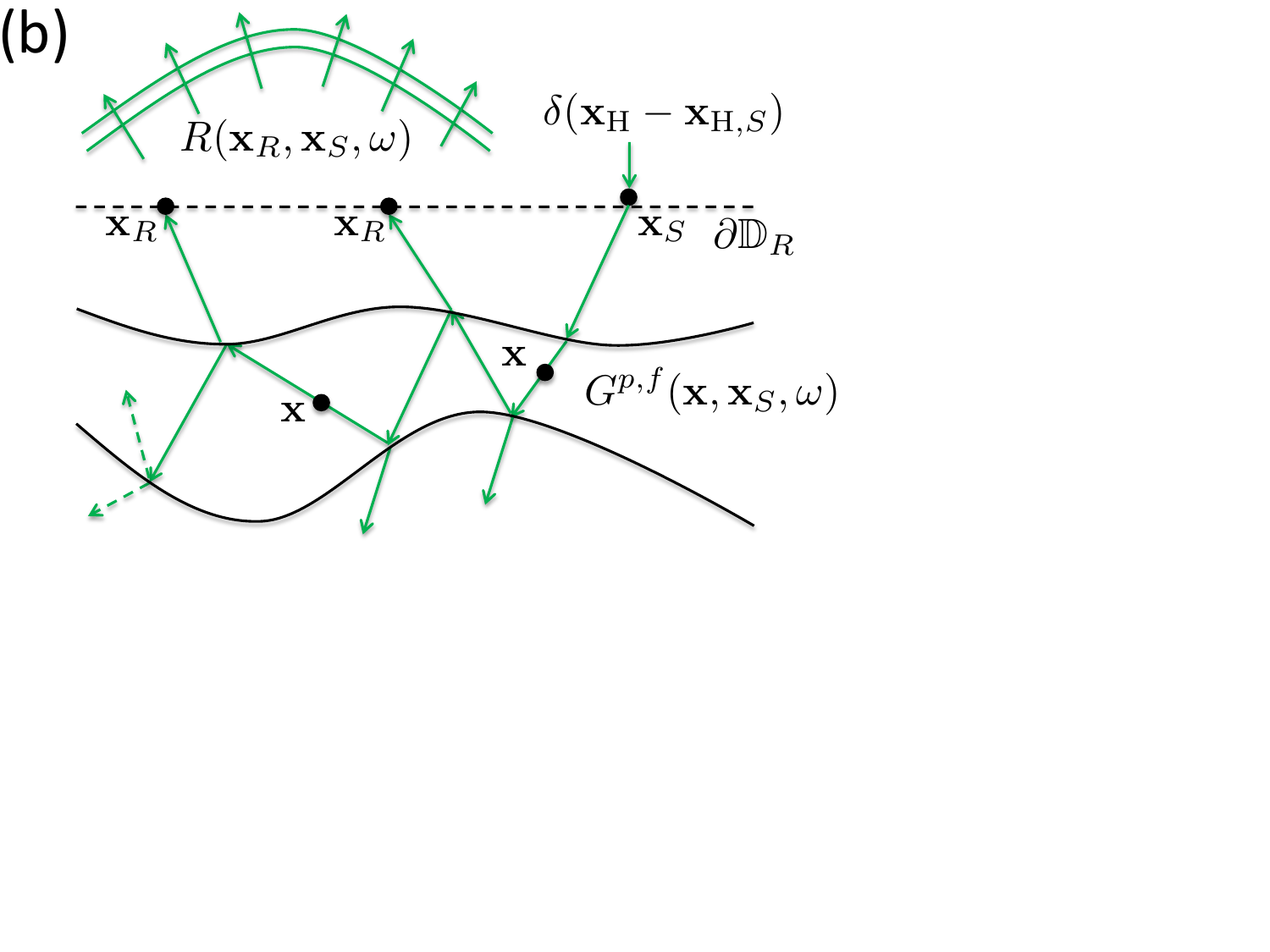}}
\vspace{-2cm}
\caption{\small Visualisation of (a) the focusing function $F({\bf x},{\bf x}_R,\omega)$ and 
(b) the Green's function $G^{p,f}({\bf x},{\bf x}_S,\omega)$ and the reflection response $R({\bf x}_R,{\bf x}_S,\omega)$.}\label{Figure2}
\end{figure}

\section{Green's matrix representations}

We define the Green's matrix ${\bf G}({\bf x},{\bf x}_S,\omega)$ as the solution of wave equation \ref{eq2.1} with a unit source at ${\bf x}_S$, hence
\begin{eqnarray}\label{eq2.1g}
\partial_3{\bf G} ={{{\mbox{\boldmath ${\cal A}$}}}}{\bf G} +{\bf I}\delta({\bf x}-{\bf x}_S).
\end{eqnarray}
Moreover, we demand that ${\bf G}$ obeys Sommerfeld's radiation condition at infinity.
We partition ${\bf G}$ as follows
\begin{eqnarray}\label{eq423}
{\bf G}({\bf x},{\bf x}_S,\omega)= \begin{pmatrix}G^{p,f}      & G^{p,q} \\
                G^{v,f} & G^{v,q}    \end{pmatrix}({\bf x},{\bf x}_S,\omega),
\end{eqnarray}
with the first and second superscript referring to the field quantity at ${\bf x}$ and the source quantity at ${\bf x}_S$, respectively.
We choose ${\bf x}_S$ at a vanishing distance above ${{\partial{\mathbb{D}}}}_R$. 
For this situation, we write for the downgoing and upgoing components of $G^{p,f}$ at ${{\partial{\mathbb{D}}}}_R$ (i.e., just below the source)
\begin{eqnarray}
2G^{p,f+}({\bf x},{\bf x}_S,\omega)|_{x_3={x_{3,R}}}&=&\delta({\bf x}_{\rm H}-{\bf x}_{{\rm H},S}),\label{eq4Gag}\\
2G^{p,f-}({\bf x}_R,{\bf x}_S,\omega)&=&R({\bf x}_R,{\bf x}_S,\omega),
\end{eqnarray}
with ${{\bf x}_{{\rm H},S}}=(x_{1,S},x_{2,S})$, and $R({\bf x}_R,{\bf x}_S,\omega)$ denoting the reflection response of the inhomogeneous medium below ${{\partial{\mathbb{D}}}}_R$, see Figure 2b.
Substitution of $G^{p,f}$ and $G^{p,f\pm}$ for $p$ and $p^\pm$ in equation \ref{eq12} gives
 \begin{eqnarray}
2G^{p,f}({\bf x},{\bf x}_S,\omega)&=&\int_{{{\partial\mathbb{D}}_R}} F({\bf x},{\bf x}_R,\omega)R({\bf x}_R,{\bf x}_S,\omega){\rm d}{\bf x}_R
+F^*({\bf x},{\bf x}_S,\omega),\label{eq339}
\end{eqnarray}
for $x_3 \ge {x_{3,R}}$. This representation (when transformed to the time domain) has a comparable form as equation 13 in \citet{Wapenaar2014GEO}.
Hence, it forms the basis for a Marchenko scheme to derive the focusing function $F$ from the reflection response $R$ and an estimate of the direct arrival of $F$. 
However, unlike in the aforementioned reference we did not assume that, inside the medium, $F$ can be decomposed into downgoing and upgoing constituents 
and that the evanescent field can be ignored. Here we only made such assumptions in the homogeneous upper half-space (including ${{\partial{\mathbb{D}}}}_R$).
Hence, the representation of equation \ref{eq339} accounts for example for refracted waves in high-velocity layers
and it remains valid in caustics. How to exploit the more general validity of 
this representation for the retrieval of the focusing function in complex cases is subject of current research. 
In particular, it needs to be investigated how to deal with the temporal overlap of the Green's function and the focusing function for refracted and evanescent waves.

The propagator matrix ${\bf W}$ can be constructed from the focusing function $F$ via equations \ref{eq424} and \ref{eq523} $-$ \ref{eq1124W}.
Assuming $F$ is obtained with the traditional Marchenko method, ${\bf W}$ inherits its relative insensitivity to inaccuracies in the subsurface model: 
its direct arrivals come from a macro model and its scattering coda from the reflection response $R$ at the surface.
Subsequently, ${\bf W}$ can be used in equation \ref{eq1330} for ``migration based  on the two-way wave equation'' \cite{Wapenaar86GP2}.
Replacing ${{\bf q}}$ by ${\bf G}$ in equation \ref{eq1330} we obtain
\begin{eqnarray}
{\bf G}({\bf x},{\bf x}_S,\omega)=\int_{{{\partial{\mathbb{D}}}}_R} {\bf W}({\bf x},{{\bf x}_R},\omega){\bf G}({\bf x}_R,{\bf x}_S,\omega){\rm d}{{\bf x}_R},\label{eq1330G}
\end{eqnarray}
for $x_3\ge x_{3,R}>x_{3,S}$. Hence, ${\bf W}$ can also be used for retrieving the complete Green's matrix between the surface and any subsurface location. 
Finally, we show that it can be used for retrieval of the homogeneous Green's matrix between two subsurface locations.
We define this matrix as ${\bf G}_{\rm h}({\bf x},{\bf x}_A,\omega)={\bf G}({\bf x},{\bf x}_A,\omega)-{\bf J}{\bf G}^*({\bf x},{\bf x}_A,\omega){\bf J}$, with ${\bf J}={\rm diag}(1,-1)$. 
Using ${\bf J}{{{\mbox{\boldmath ${\cal A}$}}}}^*{\bf J}={{{\mbox{\boldmath ${\cal A}$}}}}$, it follows that ${\bf G}_{\rm h}$ obeys equation \ref{eq2.1g} without the source term, analogous to the scalar homogeneous Green's function \cite{Oristaglio89IP}.
Replacing ${{\bf q}}$ by ${\bf G}_{\rm h}$ in equation \ref{eq1330} we obtain
\begin{eqnarray}
{\bf G}_{\rm h}({\bf x},{\bf x}_A,\omega)=\int_{{{\partial{\mathbb{D}}}}_R} {\bf W}({\bf x},{{\bf x}_R},\omega){\bf G}_{\rm h}({\bf x}_R,{\bf x}_A,\omega){\rm d}{{\bf x}_R},\label{eq1330GH}
\end{eqnarray}
where the arrangement of $x_{3,R}$, $x_{3,A}$ and $x_3$ is arbitrary (since ${\bf G}_{\rm h}$ obeys a source-free wave equation).
This generalises the scalar single-sided homogeneous Green's function representation \cite{Wapenaar2017GP2}. 

\section{Conclusions}

We have shown that the focusing function used in Marchenko imaging is intimately related to the propagator matrix. 
By deriving the focusing function directly from the propagator matrix, we circumvented up-down decomposition and did not ignore evanescent waves inside the medium. 
This may ultimately lead to more general Marchenko schemes, with the ability to accurately image steep flanks and to account for evanescent and refracted waves.
Conversely, by constructing the propagator matrix from the focusing function obtained with the traditional data-driven Marchenko method, 
the propagator matrix may be used in migration
and Green's matrix retrieval schemes, circumventing the sensitivity of the model-driven propagator matrix to the subsurface model.
Last, but not least, the matrix-vector formalism used in this paper facilitates a generalisation of the discussed relations to other wave phenomena. 

\section{Acknowledgments}

We acknowledge funding from 
 the European Research Council (ERC) under the European Union's Horizon 2020 research and innovation programme (grant agreement No: 742703).

\section{Data and materials availability}

No data have been used for this study.

\newpage
\newpage
\centerline{\Huge Captions}

\noindent
Figure 1. (a) Horizontally layered medium. (b) Symmetric propagator element $W^{p,p}(s_1,x_3,{x_{3,R}},\tau)$  (fixed $s_1$ and $x_{3,R}$), convolved with a wavelet. 
(c) Anti-symmetric propagator element $W^{p,v}(s_1,x_3,{x_{3,R}},\tau)$. (d) Focusing function $F(s_1,x_3,{x_{3,R}},\tau)$.\\

\noindent
Figure 2. Visualisation of (a) the focusing function $F ({\bf x} ,{\bf x} _R,\omega )$ and (b) the Green's function $G^{p,f}({\bf x},{\bf x}_S,\omega )$ and the reflection response $R({\bf x}_R,{\bf x}_S,\omega )$.

\end{spacing}

\end{document}